\documentclass[reprint]{elsarticle}
\usepackage{lineno,hyperref}
\modulolinenumbers[5]
\usepackage{graphicx}
\usepackage{subcaption}
\usepackage{placeins}
\usepackage{multirow}
\usepackage{amsmath}
\usepackage{soul}
\usepackage{color}
\usepackage{ulem}
\usepackage{mathptmx}

\begin{document}
	
	\begin{frontmatter}
		
		\title{Effect of ambient on the dynamics of re-deposition in the rear laser ablation of a thin film}
		
		\author[affiliation1,affiliation2]{Renjith Kumar R\corref{correspondingauthor1}}
		\cortext[correspondingauthor1]{Email:renjithkumar.cpd@gmail.com}
		
		\author[affiliation1,affiliation2]{B R Geethika}
		\author[affiliation1,affiliation3]{Nancy Verma}
		\author[affiliation1]{Vishnu Chaudhari}
		\author[affiliation1]{Janvi Dave}
		\author[affiliation1]{Hem Chandra Joshi}
		\author[affiliation1,affiliation2]{Jinto Thomas\corref{correspondingauthor2}}
		\cortext[correspondingauthor2]{Email:jinto@ipr.res.in}
		
		\address[affiliation1]{Institute For Plasma Research, Bhat, Gandhinagar, Gujarat, 382428, India}
		\address[affiliation2]{Homi Bhabha National Institute, Training School Complex, Anushaktinagar, Mumbai, 400094, India}
		\address[affiliation3]{Department of Physics, Allahabad Degree College, University of Allahabad, 211002, India}
		\begin{abstract}
			In this work, we report an innovative pump-probe based experimental set up, to study the melting, subsequent evaporation, plasma formation and redeposition in a thin film coated on a glass substrate under different ambient conditions and laser fluences. The ambient conditions restrict the expansion of the plasma plume. At high ambient pressure, plume expansion stops closer to the substrate and get re-deposited at the site of the ablation. This helps in the identification of multiple processes and their temporal evolutions during the melting, expansion and re-deposition stages. The ambient conditions affect the plasma plume formed upon ablation, thus modulating the transmission of probe laser pulses, which provides information about the plume dynamics. Further, the study offers valuable insights into the laser-based ablation of thin film coatings, which will have implications in in situ cleaning of view ports on large experimental facilities such as tokamaks and other systems e.g. coating units, pulsed laser deposition, Laser induced forward transfer, Laser surface structuring, etc. 
		\end{abstract}
		
		
	\end{frontmatter}

	\section{Introduction}\label{sec:intro}
	
	Laser surface interaction has been a subject of considerable research interest due to its manifold applications and in the fundamental understanding of the process involved. Over the past few decades, the processes of melting a surface under laser irradiation and the involved process of re-solidification and plume expansion have been extensively studied. The interaction  of high-intensity laser pulse with surface results in the creation of plasma called laser produced plasma (LPP), which by its density and temperature\cite{chichkov1996femtosecond}. Laser surface interaction has enticing applications like laser induced breakdown spectroscopy (LIBS)\cite{thomas2023review}, surface structuring\cite{radziemski2020lasers} surface modification\cite{quazi2016laser,li2006surface,qiu_advanced,xie_advan} pulsed laser deposition\cite{ kwok2016developments} etc. The laser plasma production and expansion mechanism changes according to the pulse duration of the laser\cite{chichkov1996femtosecond}. For example, the LPP by nanosecond laser is accompanied by various intricate processes\cite{harilal2022optical}, e.g. laser absorption, material excitation, ionization, thermal vaporization, shock wave propagation, plasma confinement, and subsequent atomic and molecular emissions, followed by particle condensation. To understand these mechanisms of laser surface interaction and laser produced plasma, various optical diagnostics including pump probe method are used by different research groups\cite{harilal2022optical,sweeney1976interferometric}. The time scales of phase change during nanosecond laser ablation of metals such as Cu, Au, and Ni are reported by Boneberg et al. \cite{boneberg2000nanosecond} using nanosecond time resolved reflectivity measurement. In another study, Carrasco-García et al. \cite{carrasco2015visualization} employed femtosecond pump-probe time-resolved microscopy to demonstrate the dynamics of the processes taking place during femtosecond laser matter interaction, from initial surface modification to final solidification through transient melting. Wang et al. \cite{wang1992femtosecond} suggested that, surface expansion and electron heating are manifested in reflectivity measurements for Ag and Al. From the above mentioned studies it is evident that pump probe study can
	provide an appropriate platform for investigating laser
	surface interaction. Additionally, reflectivity data in all of these studies for bulk materials were interpreted using the Drude model\cite{gamaly2013physics,ujihara1972reflectivity}. 
	
	Apart from the study of laser ablation of bulk material, a few investigations on thin films in various geometries\cite{bullock1998laser,beyer2003comparison,escobar1999thin} using spectroscopic and imaging techniques have also been performed. Mondal et al\cite{mondal2019neutral}. have previously reported neutral ion density, temperature, and plume imaging in relation to back and front ablation geometries. Study done by Thomas et al\cite{thomas2018effect}. investigated the effect of ambient pressure in plume expansion for rear ablation geometry.
	
	Laser Induced Forward Transfer (LIFT) is another aspect of thin film ablation, where a laser is used to ablate a thin film and transfer the material to an acceptor substrate close to the donor substrate\cite{morales2018laser,serra2019laser,cherepakhin_advan} . It is widely used for making photo-masks and integrated circuits, as well as forming direct film topologies, etc \cite{veiko2006laser}.
	
	Further, the study on the ablation of thin film has got a reasonable interest in cleaning the plasma facing optical components like view ports in reactor grade plasma which get contaminated due to heavy deposition of C, Fe, Cr, Ni..etc\cite{maurya2015proof}.The thin film formed by this deposition can be removed from the viewports by laser irradiation\cite{mukhin2009progress} without breaking the chamber's vacuum and causing the damage to view-ports. However, not much research is done on the fluence dependence and appropriate number of pulses needed for laser cleaning\cite{oltra2000modeling}.
	
	During the ablation process, after plasma generation, the condensed particles can be re-deposited onto the target. In the past, various groups have investigated this re-deposition. Selamantianos et al.\cite{semaltianos2008nanoparticle} have reported re-deposition during femtosecond laser ablation. Singh et al.\cite{singh2005effect} have investigated re-deposition and quantification of redeposited material around the crater. 
	
	A recent simulation study by Volkov and Lin\cite{volkov2023anomalously}. predicted an anomalous delayed re-deposition possibility of approximately 90\% of the vaporised material during nanosecond laser ablation. They suggested that a secondary shock wave directed at the target is responsible for this re-deposition. In another recent study Yong et al\cite{yong2021laser},used pump probe method to study laser absorption in a low energy nanosecond laser generated spark in atmospheric pressure air.
	
	In view of this, study of laser interaction with thin films is, therefore, appropriate and relevant for pump probe studies that can reveal surface interaction, plasma expansion, re-deposition and time evolution of these processes. In the present work we report a pump-probe experiment to study the laser interaction with an aluminium thin film. The reflected and transmitted probe pulses are investigated to obtain the dynamics of the film melting and plasma plume expansion.
	
	\section{Experimental Set-up}\label{sec:setup}
	
	\begin{figure}[hbtp]
		
		\centering
		\centering
		\includegraphics[width=1\textwidth]{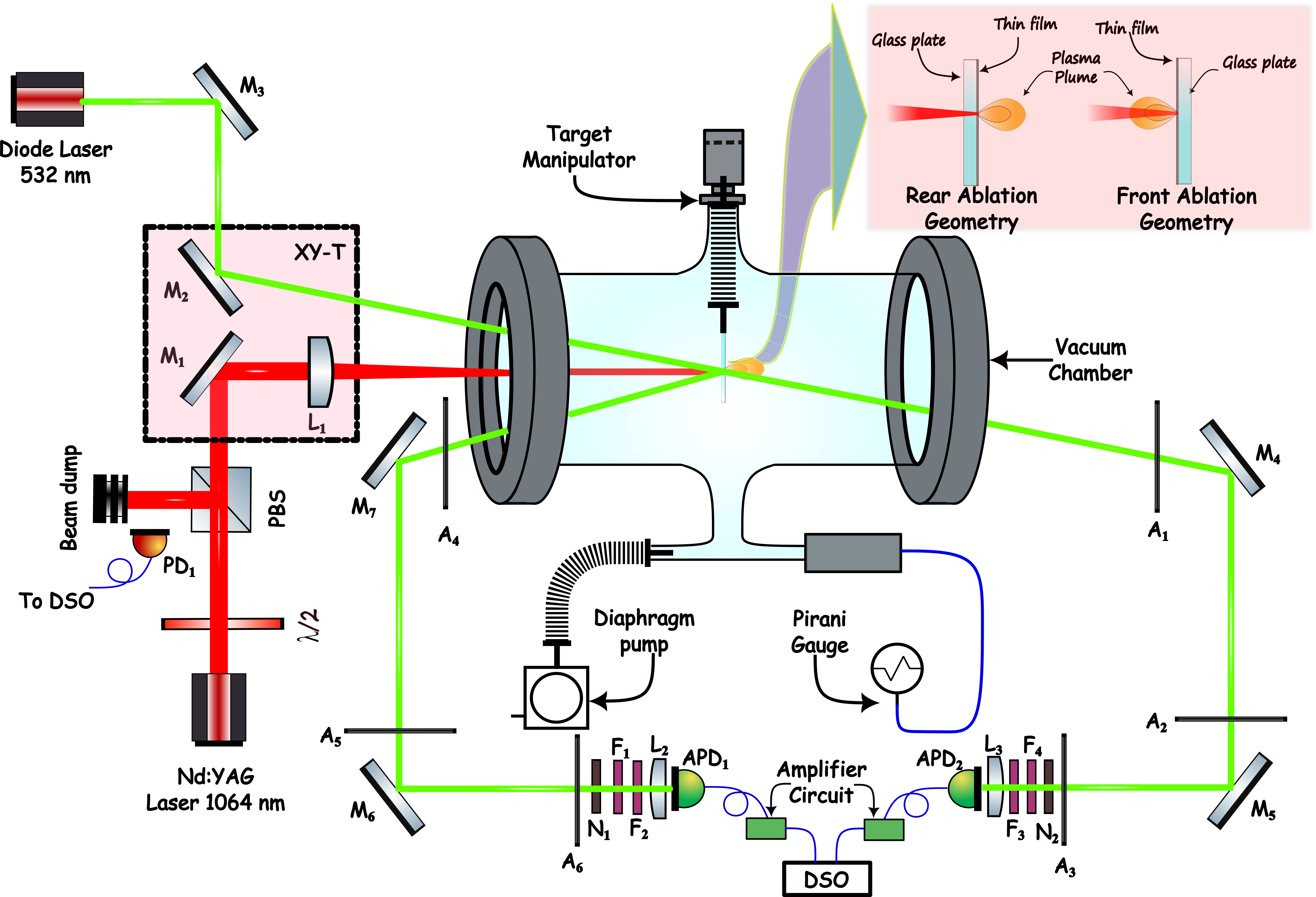}
		\caption{Experimental pump probe setup for measuring transmittance and reflectance. M-Mirror, A-Aperture, L-Lens, N-Neutral density filter. F-interference filter, APD- Avalanche photodiode, PBS-Polarising beam splitter, PD-Photodiode, $\lambda/2$-Half wave plate, XY-T-Automated XY translation stage, DSO-Digital Storage Oscilloscope.}
		\label{fig:experimental_setup}
		
	\end{figure}
	
	Figure ~\ref{fig:experimental_setup} shows schematic diagram of the experimental setup, which consists of a 10 cm diameter glass chamber having a length of 35 cm. The chamber is coupled to a diaphragm pump to evacuate it to a base pressure of 0.01 mbar. A Pirani gauge is used to monitor the pressure inside the chamber.   The sample holder for mounting the thin film is coupled to a vacuum-compatible linear motion feedthrough to vary its position after each laser shot, as shown in the figure \ref{fig:experimental_setup} depicted as the target manipulator.   
	
	A 410 nm thin aluminium thin film is coated on a fused silica substrate and used as the sample surface for investigation.  The fundamental wavelength of a Q-switched Nd:YAG laser(Powerlite 9030,Continuum) with a pulse width of 10 ns, was used for ablating the sample. Rear ablation geometry is employed in this experiment, which allows to utilise the transmitted and reflected parts of the probe laser to study the dynamics. A plano-convex lens(L1) with a focal length of 50cm is used to focus the probe laser beam on the thin film with a spot diameter $\approx$ 1mm. The fluence used in the experiment is varied from $\approx$$2 \text{ J/cm}^{2}$ to $\approx$ $7.6\text{ J/cm}^{2}$.   This is higher than the ablation threshold of Aluminium\cite{cabalin1998experimental} ($>1.01\text{ J/cm}^{2}$ )) but much lower than the damage threshold of the substrate, fused silica, \cite{cao2018wavelength} ($<200  \text{ J/cm}^{2}$  ), which ensures that substrate remain intact during the film ablation. The laser system employed for ablation is operated at a level that is significantly below the optimal parameters to decrease laser energy. The laser energy was initially adjusted in a coarse manner by varying the flash lamp voltage and the timing of the flash lamps. Subsequently, fine-tuning was achieved through the use of a half-wave plate and a polarizing beam splitter. Additionally, the experiments were conducted at a distance of approximately five meters from the laser. This results in the laser beam profile becoming slightly elliptical in shape. A continuous-wave (CW) diode-pumped solid-state (DPSS) laser with a wavelength of 532 nanometres (nm), power of 100 milliwatts (mW), and beam diameter of approximately 1 millimetre is employed as the probe laser. This is precisely aligned at the point at an angle of incidence of 15$^\text{o}$ where the pump laser is focused on the sample. Avalanche photodiode (APD S8890-02 Hamamatsu) biased for moderate gain is utilised for recording the intensity variation of probe laser due to the laser interaction with the surface. The response time of the APD with the electronics for the gain used in the experiment is shorter than 10 ns.
	The output of the APD is connected to a fast digital storage oscilloscope to record the intensity variation. A fast photodiode (PD1) is used for triggering the oscilloscope as shown in figure \ref{fig:experimental_setup}, the trigger pulse of PD1 defines the time of the incidence of the laser pulse on the sample. Bandpass and short-pass filters are used to cut down the pump laser and plasma background reaching the detector.  With the help of the target manipulator, a fresh location is made available for each experiment and the experiments are repeated multiple times to reduce statistical deviations in result due to laser energy fluctuations and surface non-uniformity.  The decrease in laser energy slightly distorts the beam shape. The morphology of the ablated surface was examined with a scanning electron microscope(SEM). Micro-Raman spectra of the ablated spots using 100x objective lens were acquired using a Horiba make micro-Raman spectrometer using 532 nm as the laser wavelength.
	
	\section{Results and discussion}\label{sec:results}
	
\begin{figure}[hbtp]
		\centering
		\includegraphics[width=1\textwidth]{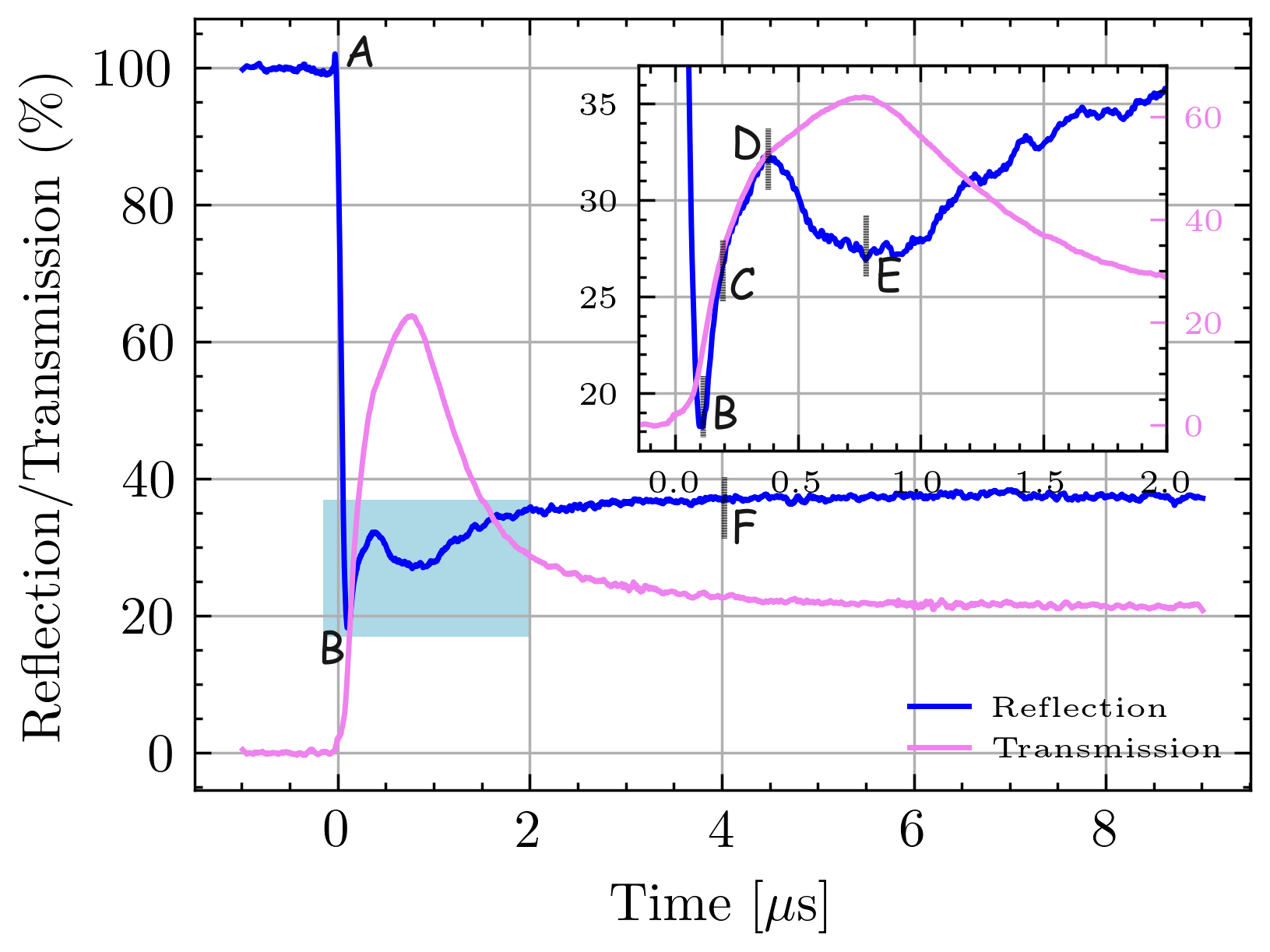}
		\caption{Temporal evolution of transmission and reflection at atmospheric pressure at a fluence 5.0$\text{ J/cm}^{2}$. Each traces are the average of five experiments under the same experimental conditions. The inset is a magnified portion of the reflectance overlaid with transmission.}
		\label{fig:ba_refl_and_trans_p1_20mj}
	\end{figure}

	Figure \ref{fig:ba_refl_and_trans_p1_20mj} shows the time evolution of the transmitted and reflected intensities of the probe beam for the pump laser fluence of  5.0 $\text{ J/cm}^{2}$. The absolute value of reflection is estimated considering the initial reflected power as 100 \%, and the reflection from the location where the film is completely ablated as the zero reflection. For transmission measurements, the location where the film has been completely ablated is regarded as having 100\% transmission.
	To reduce statistical errors, each data point in the figure represents the average of multiple experiments conducted under similar conditions. The sharp peak at the location marked as A in figure \ref{fig:ba_refl_and_trans_p1_20mj} for the reflected signal is the pump laser pulse pick-up by the detector at the time of incidence which persists despite the introduction of band-pass and edge filters. The inset in the figure \ref{fig:ba_refl_and_trans_p1_20mj} is the zoomed portion of the early time evolutions of transmission and reflection.  The reflected portion of the probe beam shows a drastic decrease in intensity initially which reaches 20\%  within 100 ns (point marked as B in inset). After reaching the minimum value, the reflectivity exhibits a unique pattern with a small crest around 350 ns(D) followed by a trough at 800 ns(E). After that, the reflectivity gradually increases and remains almost constant from ~4 $\mu s$ onwards.
	
	In contrast, the probe laser transmission increases at a significantly slower rate than the decrease in reflection. The transmission reaches a maximum of approximately 65\% at around 800 ns, which suggests that the film has undergone substantial ablation due to the pump laser. Subsequently, a notable decline in transmission is observed, with a constant level of 25\% being maintained from approximately 4 $\mu s$ onwards. Indeed, the experimental configuration, which involved the reflection and transmission of the probe, was utilized to provide comprehensive insights into the melting of the film and the absorption of plasma. However, due to the rapid evolution of the material's physical properties during laser interaction and the potential for reflection from the plasma at critical density, it is rather challenging to achieve a clear distinction during the initial stages of plasma formation.

	\par
	It is evident from the inset that the decrease in reflection occurs at a faster rate than the increase in transmission, with the former reaching minimum before the latter become maximum.
	The transmitted and reflected parts of the beam can provide information about the effective temporal characteristics of the melting and subsequent evolution of the plasma plume. On the other hand, transmitted beam can exhibit combined effect of film melting and the interaction of it with the laser plasma formed for a duration of 100ns. For example, the transmission initially reaches nearly 65\% and then falls back to nearly 25\%, indicating partial ablation of the film on ablation.
	
\begin{figure}
		\centering	
		\begin{subfigure}[b]{0.45\textwidth}
			\centering
			\includegraphics[width=1\textwidth]{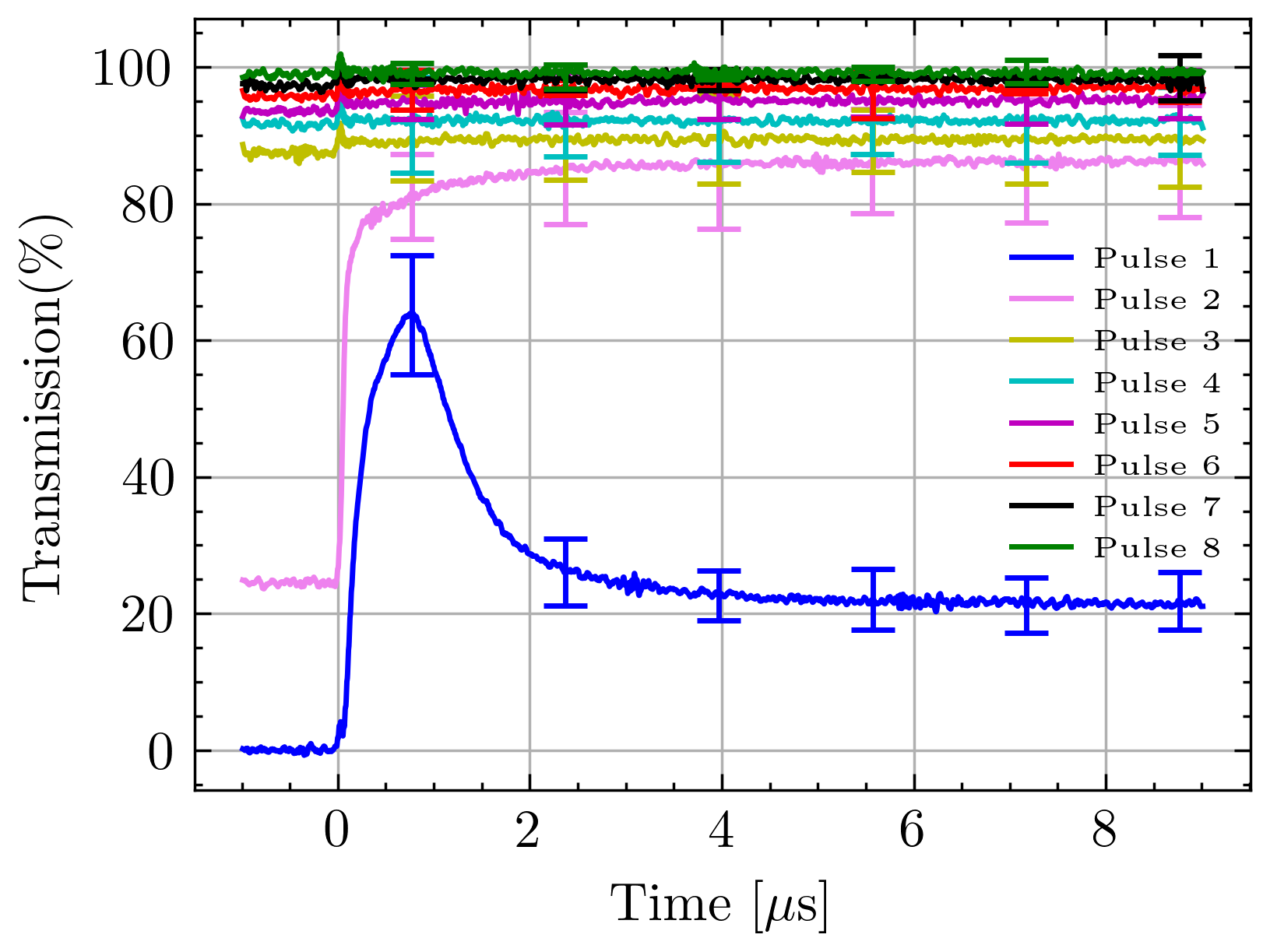}
			\caption{}
			\label{fig:ba_tdata_var_Pul_20mJ}
		\end{subfigure}
		\hfill
		\begin{subfigure}[b]{0.45\textwidth}
			\centering
			\includegraphics[width=1\textwidth]{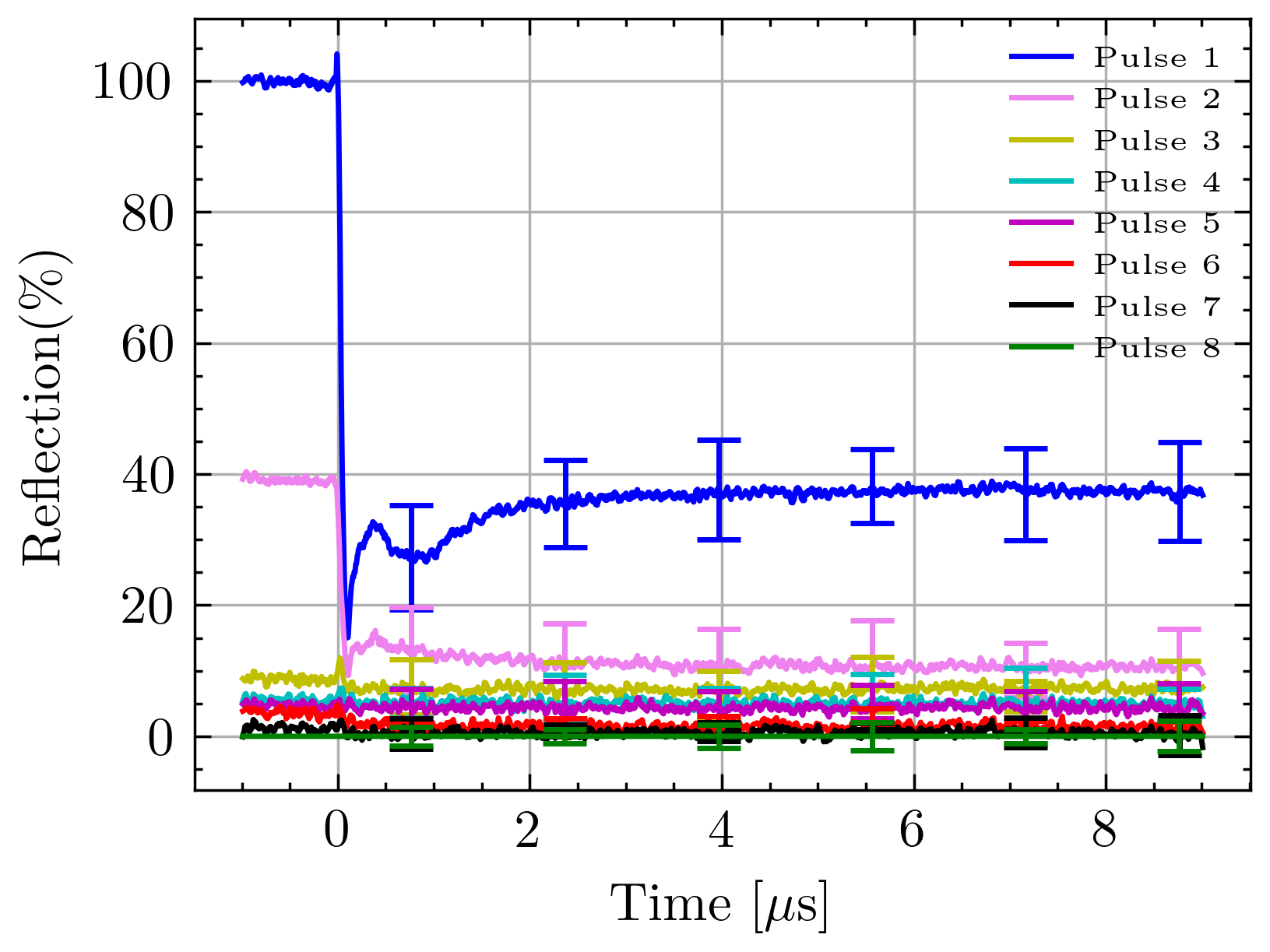}
			\caption{}
			\label{fig:ba_rdata_var_Pul_20mJ}
		\end{subfigure}
		\caption{Temporal evolution of transmission and reflection for rear ablation at a laser fluence 5.0$\text{ J/cm}^{2}$ for different pulses at same spot at a background pressure of 1000 mbar.}
	\end{figure}

	To confirm this, the film is ablated again at the same spot and the transmission is recorded.  Figure \ref{fig:ba_tdata_var_Pul_20mJ} displays the evolution of transmission through the same spot, when it is ablated with a series of laser pulses. As stated earlier, the graph plotted is the data averaged over multiple experiments at the same experimental conditions and the error bars are the standard deviation.  To enhance the visibility of data, the error bars are kept only for few instances. From figure \ref{fig:ba_tdata_var_Pul_20mJ}, it can be observed that prior to the ablation with the second laser pulse,  transmission level remains nearly constant at the saturation point established by the first pulse. This indicates that there are no significant changes in transmission after a few microseconds.  The temporal evolution of the transmission for the second ablation on the same point is significantly different to that of first pulse, as can be seen in figure \ref{fig:ba_tdata_var_Pul_20mJ}. The transmission for the second pulse increases at a faster rate initially, followed by a slow increase to saturation from 3 microseconds onwards and reaches a transmission of around 85 \%. It is important to note that the transmissions does not show any decrease from its peak value as seen for the first pulse. The subsequent ablations do not show any drastic enhancement in transmission and by the $7^{th}$ pulse, the film is completely ablated. These results confirm that the first pulse ablates the film partially. The interesting observation of decrease in transmission from a peak value can be an indication of the re-deposition of the ablated material back to the substrate. However, the subsequent pulses do not show such re-deposition as the ablated mass is rather small to effectively decreases the transmission at a detectable range. In a recent simulation study by Volkov and Lin.\cite{volkov2023anomalously} it was observed that a substantial portion of the ablated material is re-deposited at the sample. It can be noted that the present work comprehensively demonstrates the re-deposition experimentally.
	
	Figure ~\ref{fig:ba_rdata_var_Pul_20mJ} illustrates the evolution of reflection from the same location for successive pulses.  The reflection of the probe laser for the second pulse decreases rapidly however, it does not show further increase in reflection. Instead, the reflection slowly decreases and rather stays constant from 2 microseconds onwards.  This observation can be correlated with the transmission for the second pulse at the same location.  The successive ablations on the same spot decrease the reflectivity slightly but no drastic decrease is seen further and the film gets almost completely ablated by the sixth pulse.

\begin{figure}[hbtp]
		\centering	
		\begin{subfigure}[b]{0.45\textwidth}
			\centering
			\includegraphics[width=1\textwidth]{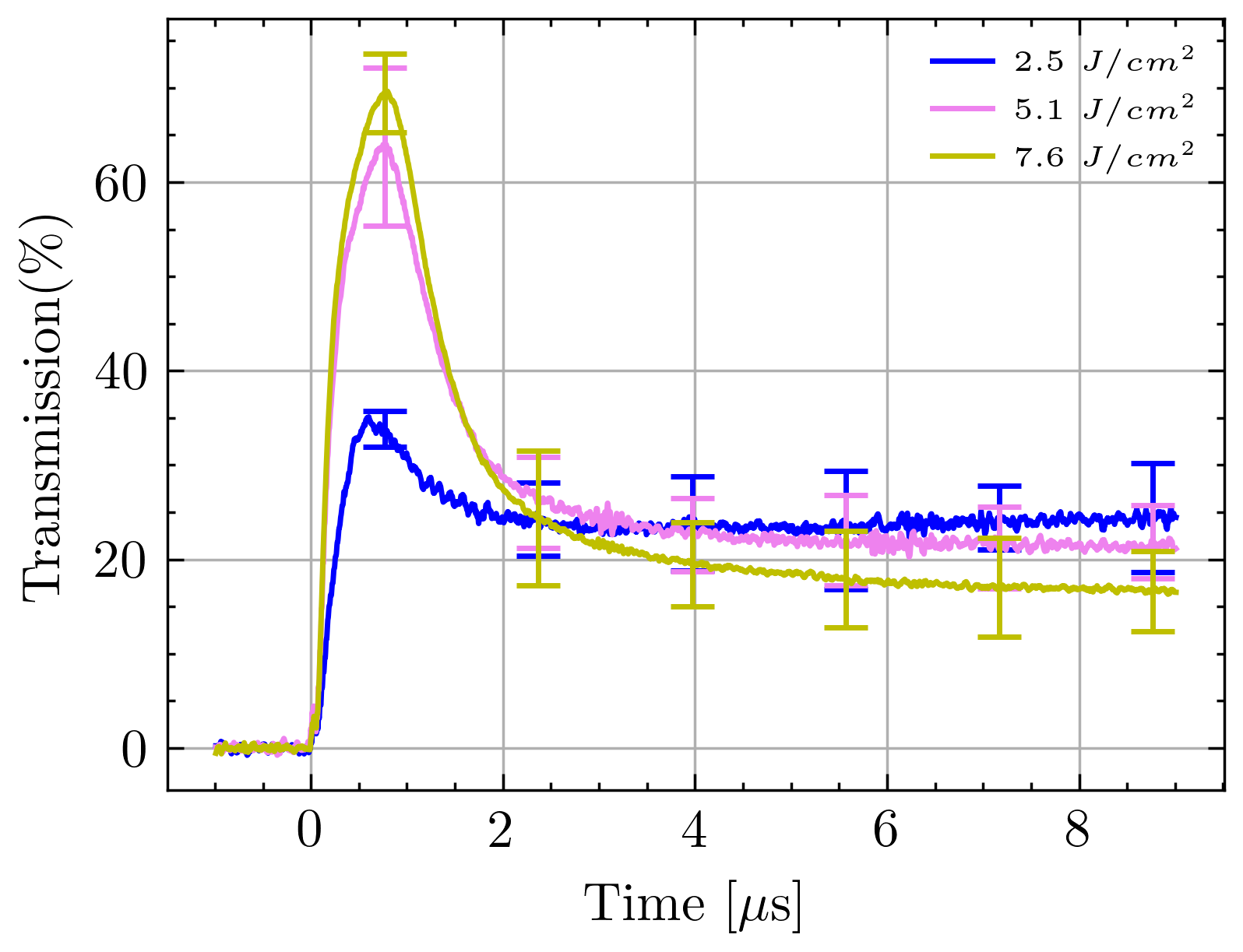}
			\caption{}
			\label{fig:ba_trans_1000mbar_var_F}
		\end{subfigure}
		\hfill
		\begin{subfigure}[b]{0.45\textwidth}
			\centering
			\includegraphics[width=1\textwidth]{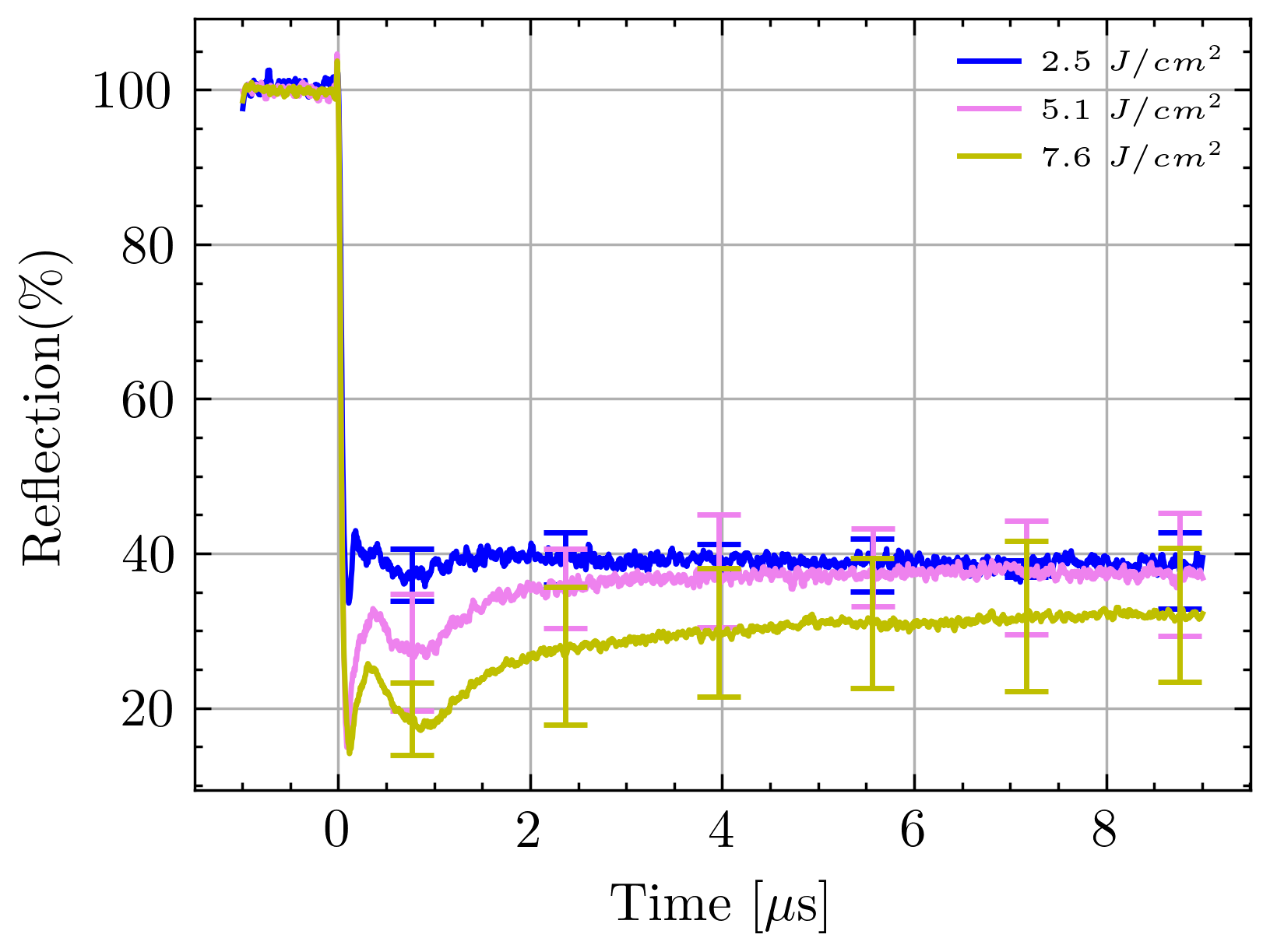}
			\caption{}
			\label{fig:ba_refl_1000mbar_var_F}
		\end{subfigure}
		\caption{Temporal evolution of transmission and reflection for different fluences at background pressure of 1000 mbar}
	\end{figure}

	As the evolution of reflection and transmission at the ablation point shows incomplete melting, it would be interesting to investigate the effect of laser fluence on the film melting in this geometry. Figure 
	\ref{fig:ba_trans_1000mbar_var_F} shows the impact of laser fluence on probe laser transmission for three different pump laser fluences. At a pump laser fluence of $2.5 \text{ J/cm}^{2}$, the peak of transmission reaches up to 35 \% and then falls to nearly 25 \%. However, in the case of higher fluences $5.1 $ and  $7.6 \text{ J/cm}^{2}$, the peak value increases beyond 60 \% and later comes down to about 20 \%. It is worth mentioning that although the peak transmission at around 800 ns shows a direct dependence on the laser fluence, the final transmission values do not show that much difference as the average values of different fluences fall within the error bar.

	The effect of variation in fluence is recorded for the reflection also. Figure ~ \ref{fig:ba_refl_1000mbar_var_F} shows the evolution of the reflectivity of the sample for different laser fluences. The figure shows that the pattern of temporal evolution is similar for all the laser fluences. Further, the reflectivity at the spot of ablation decreases with pump laser fluence. Moreover, both the reflection and transmission from the film generally depend on its thickness\cite{lugolole2015effect}. Temperature of the film is an another factor which determines the reflection\cite{boneberg2000nanosecond,volkov2023anomalously,yong2021laser}. The absorption of the probe laser by the plasma formed can substantially decrease the transmission and can affect reflection also.  It is expected that as the film thickness increases, the transmission will decrease and the reflection will increase. However, this is not clearly evident in our experimental data.

\begin{figure}[hbtp]
		\centering	
		\begin{subfigure}[b]{0.45\textwidth}
			\centering
			\includegraphics[width=1\textwidth]{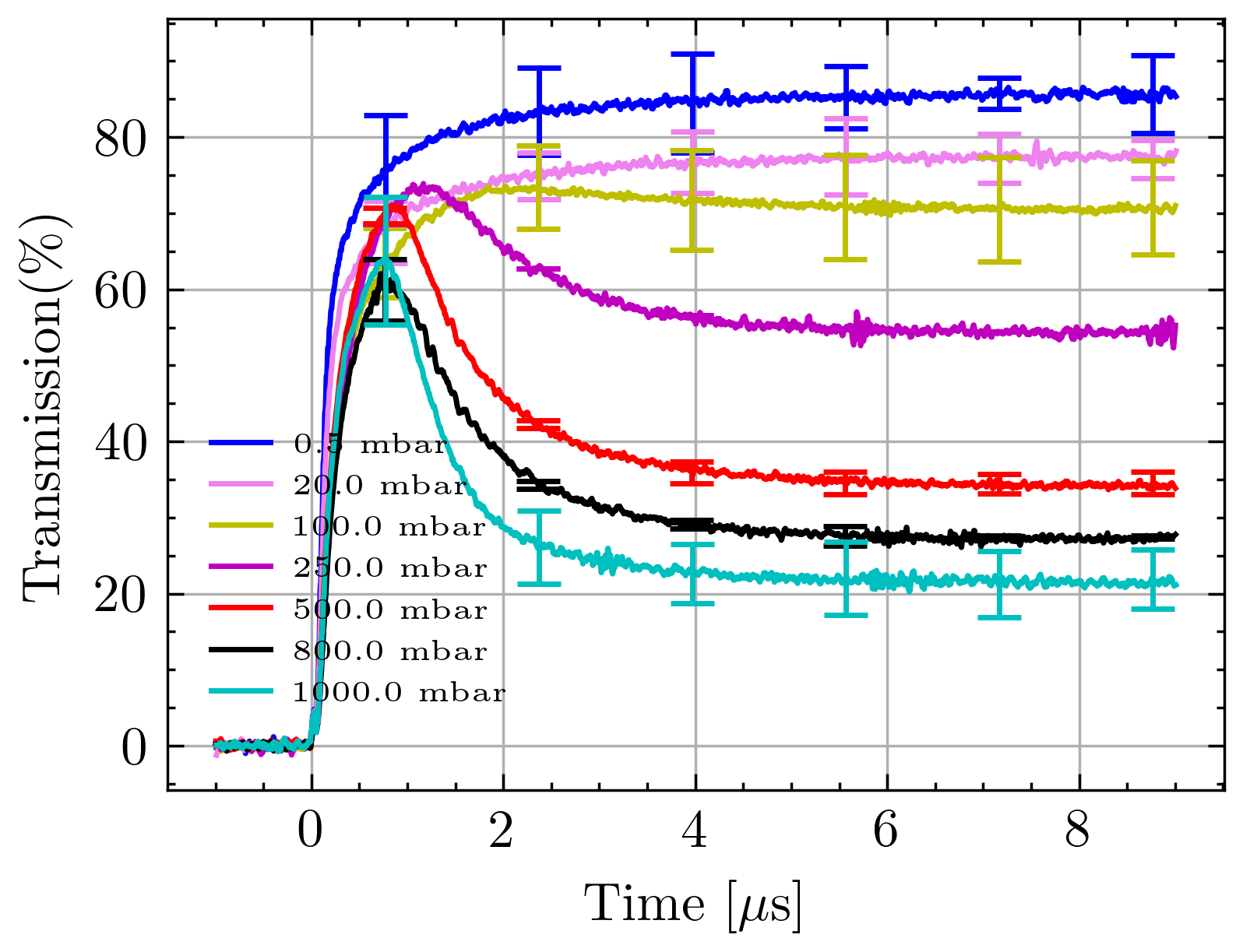}
			\caption{}
			\label{fig:ba_trans_p1_var_press_20mj}
		\end{subfigure}
		\begin{subfigure}[b]{0.45\textwidth}
			\centering
			\includegraphics[width=1\textwidth]{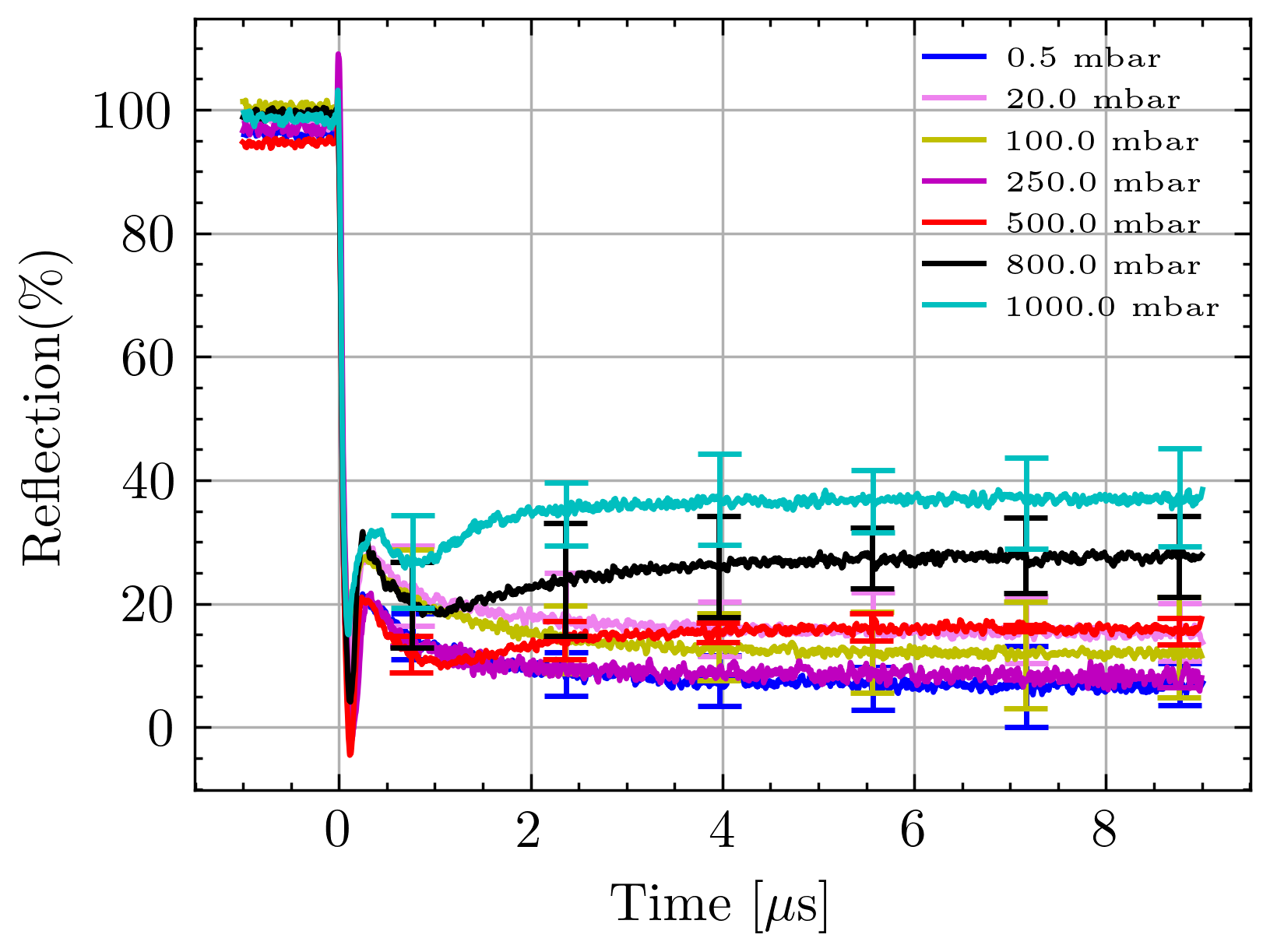}
			\caption{}
			\label{fig:ba_refl_p1_var_press_20mj}
		\end{subfigure}
		\caption{Evolution of transmission and reflection for pulse 1 at a fluence of 5.0 $\text{ J/cm}^{2}$  for different background pressures.}
	\end{figure}

	As mentioned, The temporal evolution of transmission and reflection at the ablation spot can provide insights into the potential re-deposition of the ablated material onto the substrate. At atmospheric pressure, the background medium can obstruct the plasma plume's propagation, thereby limiting its expansion. To investigate the influence of background pressure on the film ablation and subsequent probing of reflection and transmission by the probe laser, experiments were conducted under various background pressures.
	Figure \ref{fig:ba_trans_p1_var_press_20mj} illustrates the impact of background pressure on the evolution of the transmission of the probe at a pump laser fluence of 5.0 $ \text{ J/cm}^{2}$. The results demonstrate that the temporal evolution of the transmission is significantly affected by the background pressure. 
	As the background pressure decreases, the width of the transmission peak around 800 ns increases. The appearance of this peak can be attributed to the onset of redeposition. Up to 800 ns, the transmission increases due to the melting and expansion of the plasma plume. However, as the background pressure increases, redeposition starts taking place, resulting in a decrease in transmission and the formation of a peak at approximately 1 $\mu$s. 
	At approximately 100 mbar, the transmission no longer exhibits the decrease observed at higher pressures. 
	This suggests that the decrease in transmission following the peak at 800 ns is attributable to the background medium. It seems reasonable to hypothesize that the plasma plume is obstructed by the background gas, which then undergoes thermalization. This allows the plume material to be redeposited at the site of ablation, resulting in a notable decrease in transmission. Further, as the background pressure decreases, the stopping distance of the plasma plume is expected to increase, thereby reducing the extent of redeposition.
	For pressure ranges below 250 mbar, the transmission increases and reaches a maximum value after 2$\mu s$, suggesting the ablation of the surface without re-deposition. In fact, as the background pressure decreases, more ablation of the film occurs, resulting in higher transmission, as can be seen in the figure \ref{fig:ba_trans_p1_var_press_20mj}. 
	
	The temporal evolution of reflectance for different background pressures is also investigated using the reflected probe beam.  Figure~ \ref{fig:ba_refl_p1_var_press_20mj} shows the effect of background pressure on the reflectance while the pump laser fluence is 5.0 $\text{ J/cm}^{2}$. The reflectance decreases significantly with the decrease in background pressure, a trend that complements the transmission data.  The temporal evolution exhibits a similar trend up to 500 ns, regardless of the background pressure. However, the evolution after 500 ns shows a clear dependence on background pressure. For higher background pressures(from 500 mbar onwards), the evolution shows a minimum around 1 $\mu s$ and then increases and saturates within 2-3 microseconds. But at lower pressures, the reflectivity falls from 20-35\% at 500 ns and does not show any increase thereafter. In the absence of any observable re-deposition at the lower background pressure, it is feasible to arrive at an estimation of the extent of ablation. At a background pressure of 0.5 mbar, the transmission was found to be nearly 85\%, with a reflection of 15\% (see figure~\ref{fig:ba_trans_p1_var_press_20mj}). As previously reported by Lugolo et al \cite{lugolole2015effect}, a 15\% reflection corresponds to a film thickness of a few tens of nm, indicating that the majority of the film (90-95\%) was ablated at 5.0 $\text{ J/cm}^{2}$ when the background pressure was 0.5 mbar. 
	
\begin{figure}
		\centering
		\centering
		\includegraphics[width=1\textwidth]{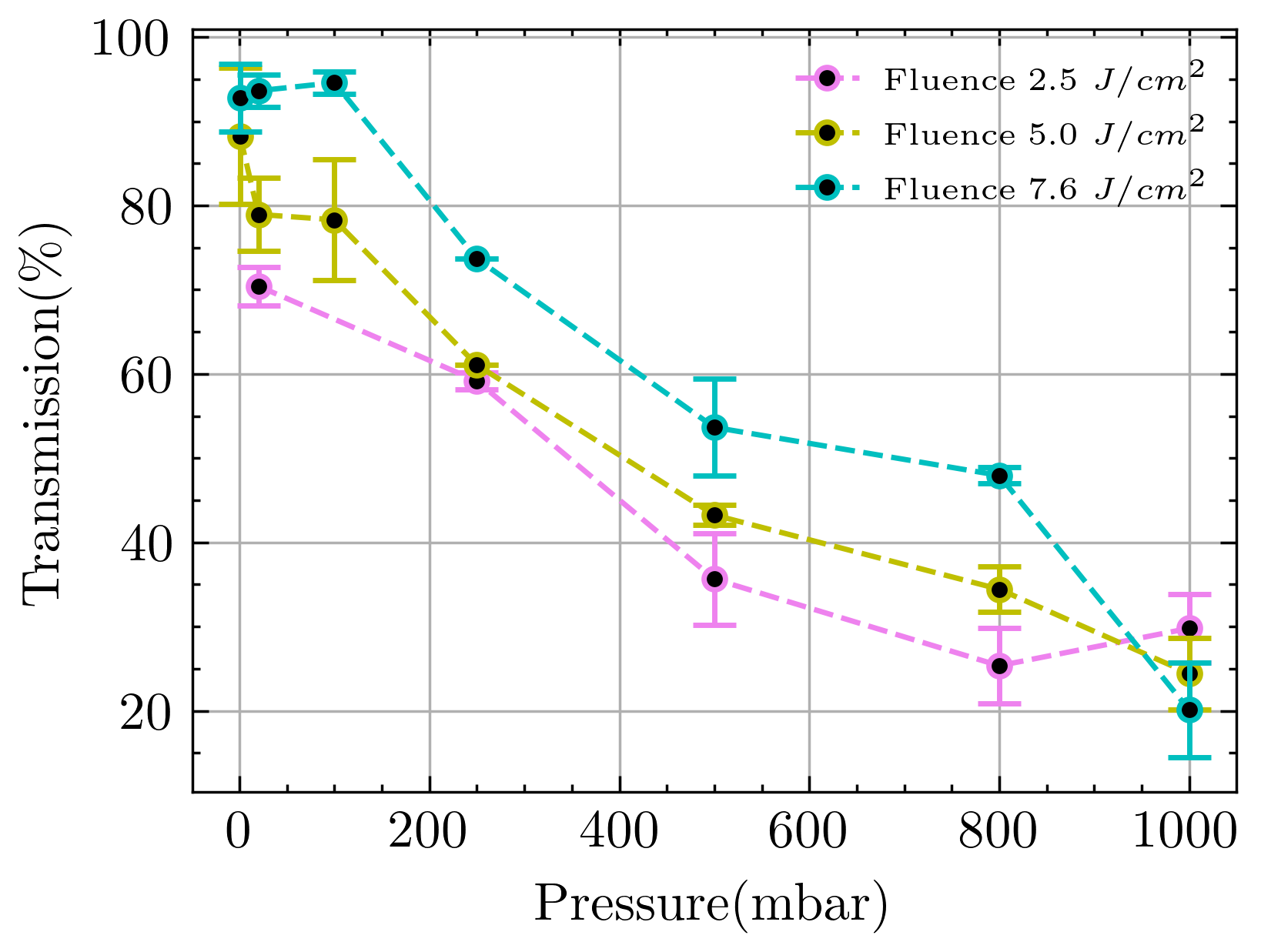}
		\caption{Change in transmission after re-deposition of the material for different pressures and fluences.}
		\label{fig:tra_decre_var_pre}
	\end{figure}

	Figure \ref{fig:tra_decre_var_pre} presents the transmission of the probe laser from the ablated spot following the first laser pulse, measured at different background pressures and laser fluences. Figure \ref{fig:tra_decre_var_pre} clearly demonstrates that the transmission decreases as the background pressure increases for all laser fluences. At atmospheric pressure, the transmission and subsequently re-deposition is so pronounced that the laser fluence has minimal impact on the transmission. These observations conclusively prove that the ablated materials is re-deposited depending on the background pressures.
	

	\begin{figure}[hbtp]
		\centering
		\begin{subfigure}[b]{0.45\textwidth}
			\centering
			\includegraphics[width=1\textwidth]{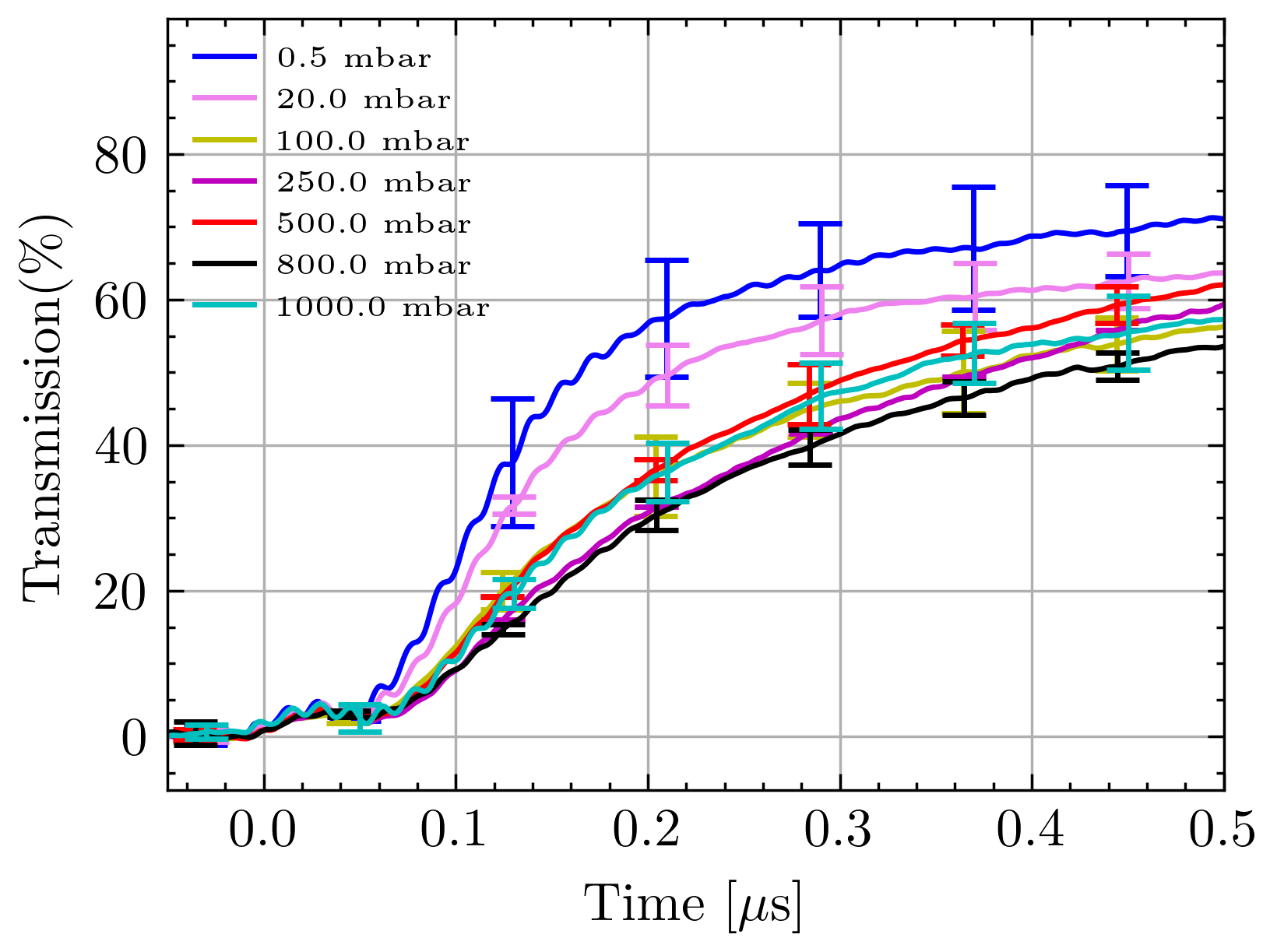}
			\caption{}
			\label{fig:ba_tdata_P1_20mJ_zoom}
		\end{subfigure}
		\hfill
		\begin{subfigure}[b]{0.45\textwidth}
			\centering
			\includegraphics[width=1\textwidth]{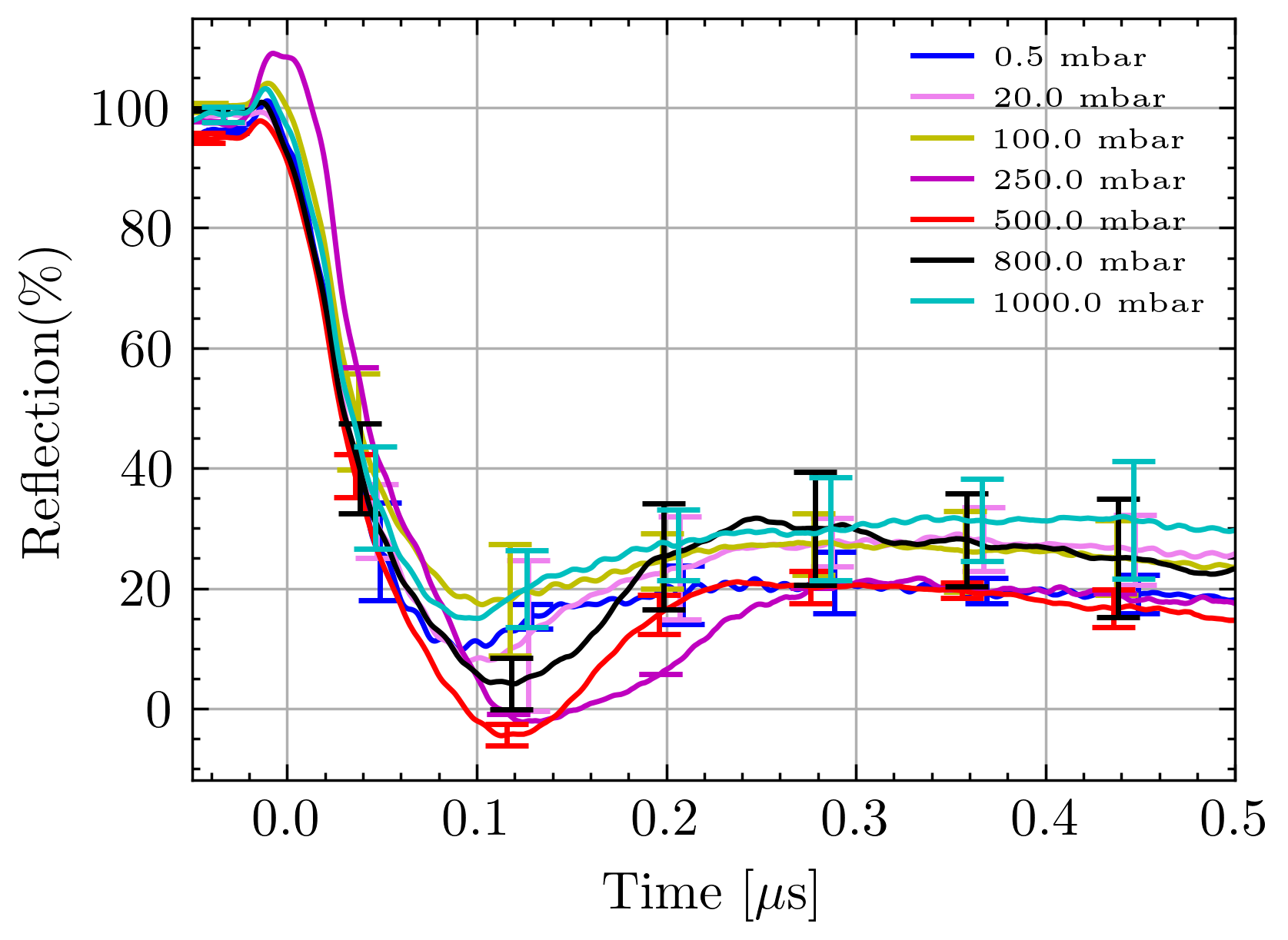}
			\caption{}
			\label{fig:ba_rdata_P1_20mJ_zoom}
		\end{subfigure}
		\caption{Evolution of transmission and reflection for different pressures for a duration of 500ns at a fluence of 5.0 $\text{ J/cm}^{2}$}
	\end{figure}

	Figure \ref{fig:ba_tdata_P1_20mJ_zoom} and \ref{fig:ba_rdata_P1_20mJ_zoom} show the early time evolutions of reflection and transmission for varying background pressures. The transmission and reflection data up to 60 ns do not show any difference in evolution for all the pressures. However, the transmission data beyond 60 ns show a faster increase in transmission as the background pressure decreases. As the melting and ablation processes of the film are not strongly dependent on the background pressure, the rapid increase in transmission can be attributed to the variation of plasma density based on the background pressure. The background pressure confines the plasma and, in turn, increases the plasma density. Further the enhanced plasma density at higher background pressures can result in increased attenuation of the probe beam by reflection or absorption of the laser photons by the plasma itself\cite{dawson1969optical}. As the background pressure is decreased, the plasma density falls more rapidly, resulting in smaller absorption and higher transmission of the probe.	
	
	The initial dynamics of the reflection signal up to 200 ns (sharp fall followed by an increase) appears to be the same for all background pressures and thus needs to be considered as intrinsic to the film.	
	Boneberg et al\cite{boneberg2000nanosecond}. observed a decrease in reflectivity as the temperature of the surface increased during laser annealing. It was observed that the reflectivity of the film decreased significantly upon laser irradiation,  the extent of this decrease being proportional to the laser fluence. Additionally, an increase in reflectivity was observed within 100 to 200 ns after laser irradiation, similar to the results observed in the present experiment. This increase in reflectivity was attributed to a the decrease in the surface temperature. As previously observed, the ablation process initiated by the first laser pulse is incomplete, resulting in the formation of a thin layer of film on the glass slide even at low background pressure. This suggests that the portion of the film that is not melted may be responsible for the observed dynamics of the reflection signal. 
	
	Thus, the overall behaviour exhibited in reflectivity can be due to the combined effect of multiple processes. Initially, upon the increase in surface temperature and melting process, reflectivity can decrease substantially\cite{ujihara1972reflectivity,Semmar}. Simulated results of reflectivity at a similar laser fluences\cite{marla2014models} show that the reflectivity decreases to nearly 50\%. However, in our experiment the reflectivity decreases to a value as low as 10\% for a duration of 100ns. This can be attributed to the expansion of the molten layer and the possible inverse bremsstrahlung (IB) absorption from the plasma plume, as evidenced by the Yong et. al.\cite{yong2021laser} and others\cite{harilal1997electron,offenberger1972plasma}. These combined effects may result in a substantial decrease in reflectivity. As time progresses, the temperature of the molten layer in contact with the substrate decreases and it re-solidifies, resulting in an increase in the reflection \cite{boneberg2000nanosecond}. The portions B to E in Figure \ref{fig:ba_refl_and_trans_p1_20mj} demonstrates the combined effect of the expansion of the plume and the re-solidification of the molten layer to the substrate. Initially, the enhancement in reflection from the re-solidification is higher than the decrease in reflection due to expansion. Subsequently, as the re-solidification process almost reaches completion, further decrease due to the expansion of the molten layer occurs. The dynamics of reflectivity evolution up to this point is consistent for all background pressures, as illustrated in Figure \ref{fig:ba_rdata_P1_20mJ_zoom}. 
	
	The evolution of the plasma plume or molten layer is significantly influenced by the background pressure \cite{thomas2018effect, farid2014emission, thomas2019pulse}. The drag force exerted by the background gas species on the plume results in its deceleration and eventual cessation of movement at the stopping distance. At higher pressures, the stopping distance may be of the order of a few millimeters\cite{sharma2007anisotropic,harilal2003internal}. Jeong et al\cite{Jeong_1999}, In an experiment with comparable laser fluence and background pressure with that of the present experiment, demonstrated that the vapour plume remained in contact with the sample for more than 5 $\mu s$ after the laser ablation. Further, Arora et al \cite{Garima_JAAS} observed that the ablated species move closer to the target as the background pressure increases. Mondal et al.\cite{mondal2019neutral} demonstrated that for back ablation,  plasma plume remains much closer to the target as compared to front ablation. Considering all these possibilities and as the plume species have still higher temperature compared to the background, it can spread in all directions through a diffusion like process resulting in a portion of it being re-deposited on the ablation spot. This further enhances the reflection and decreases the transmission, as illustrated in figure \ref{fig:ba_refl_and_trans_p1_20mj}. However at lower background pressures, re-deposition is negligible due to the longer stopping distance, which limits diffusion back to the substrate.

	Similarly, the transmission behaviour can be understood as a consequence of transmission through the plasma plume, which is by melting and evaporation, followed by the redeposition of the film at a later stage. The heat diffusion length in aluminium is 0.9 micrometers and the melting of the film occurs within a few nanoseconds\cite{boneberg2000nanosecond, schultze1991blow}, transmission remains low as the light travels through the high-density plasma plume. It is also likely that the plasma plume can also absorb laser light through inverse bremsstrahlung (IB), as discussed in the paper by Yong et al \cite{yong2021laser}. However, inverse bremsstrahlung absorption is a density and temperature-dependent process\cite{mora1982theoretical,dawson1969optical,offenberger1972plasma,harilal1997electron}, and thus is linked to the background pressure that confines the plasma plume. Consequently, the slower rise in transmission intensity compared to the fast fall in the intensity of reflection may be to IB absorption of the plasma plume shorter than 500 ns. However at this fluence IB process maynot have a significant controbution at loger times as the density is expected to be not sufficient for it. As the plasma density decreases, the transmission reaches its maximum value. 
	
	Further, as the plume stops due to the drag of the background, diffusion of the species can occur in all directions resulting in the re-deposition, as previously described. This will result in a decrease in transmission, as can be seen in the figure \ref{fig:ba_trans_p1_var_press_20mj}. In addition to these processes occurring over a duration of 400-500 ns, redeposition observed beyond this timescale may be correlated with drag model as described below. 

	The time dependent plasma plume front location $R(t)$ as per the drag model \cite{sharma2007anisotropic, thomas2018effect,harilal2003internal} is defined as  
	
	\begin{equation}
		R(t)=R_0(1-e^{-\beta t})
	\end{equation}
	where $R_0$ is the stopping distance and $v_0$ is the plume expansion velocity , $\beta $ is the drag coefficient ($\beta R_0 = v_0$). At higher pressure, the stopping distance is significantly reduced due to the aforementioned drag, as reported by numerous research groups\cite{sharma2007anisotropic,harilal2003internal,sankar2018ion,sharma2005plume}.
	For example, an aluminium plasma at 100 mbar background pressure and 331 $\text{ J/cm}^{2}$ laser fluence has an estimated stopping distance of 2.7 mm, as reported in the study by Sharma et al.  \cite{sharma2007anisotropic}. In another study by Sankar et al. (2018)\cite{sankar2018ion}, a fluence of $56 \text{ J/cm}^{2}$and atmospheric pressure resulted in a reported stopping distance of 0.5 mm. It can be extrapolated from these observations that for pressure above 250 mbar and at a significantly lower laser fluence, the stopping distance should be significantly lower than 2.5mm.
	
	If we assume that the plume is expanding to the background and stops at a distance $R_0$ from the sample. Then the ablated mass diffuses in all direction and reaches back to the ablated spot having radius r.  The percentage of redeposited mass $(\eta)$ back to the ablation spot on the target surface can be represented as,
	
	\begin{equation}
		\eta = \left( \frac{r^2}{4R_0^2}\right) \times 100
	\end{equation}
	As reported by Sharma et al. \cite{sharma2007anisotropic}, at a background pressure of 100 mbar, if we consider the stopping distance to be 2.5 mm and a crater radius of 0.5 mm, the resulting $\eta$ to be $\approx1.0 \%$, which is unlikely to be detected in the present experiment. However, if we consider the stopping distance (front position $\approx$1.4 mm) as reported by Sankar et al. \cite{sankar2018ion}, for the atmospheric pressure, the resulting value for $\eta$ to be $\approx13.0 \%$, which is significant. Furthermore, the laser fluence used in the present experiment is considerably lower than that of the aforementioned cases, and thus the amount of re-deposition can be expected to be higher, and appears that the drag model can correlates reasonably well with concept of re-deposition

	Further, a recent simulation by Volkov and Lin\cite{volkov2023anomalously} shows that there is a possibility of delayed re-deposition of $\approx$90\% vapourised material. However, it can be noted that this study has done for front ablation at a comparable laser fluence of the present work(1.5 to 4 $\text{ J/cm}^{2}$). 
	
	Moreover, it is pertinent to note that the rates of decrease in transmission and concurrent increase in reflection following the peak in transmission at approximately 800 nanoseconds appears to be the same. To substantiate this, the transmission and reflection data from 900 ns are fitted with the following relationship.

	\begin{equation}
		I(t)=I_0(1\pm e^{- \alpha t})
		\label{eq:fit_equation}
	\end{equation}
	
	where $I_0$ is the min/maximum intensities and $\alpha$ is the exponent for increase or decrease in intensity.
	The values of $\alpha$  for reflection and transmission fits in figure \ref{fig:ba_refl_and_trans_p1_20mj} are $1.355 \pm 0.003 \times10^6 s^{-1}$ and $1.544 \pm 0.007 \times10^6 s^{-1}$ respectively. Moreover, alpha increases with background pressure from a value of $0.542 \pm 0.002 \times10^6 s^{-1}$ at 250 mbar to $1.544 \pm 0.007 \times10^6 s^{-1}$ at the atmosphere. The small differences in alpha for reflection and transmission may be due to the diffused reflections from the re-deposited film. This exponential fit suggests that the re-deposition may be occur due to the diffusion of ablated mass.

	\begin{figure}[hbtp]
		\centering
		\includegraphics[width=1\textwidth]{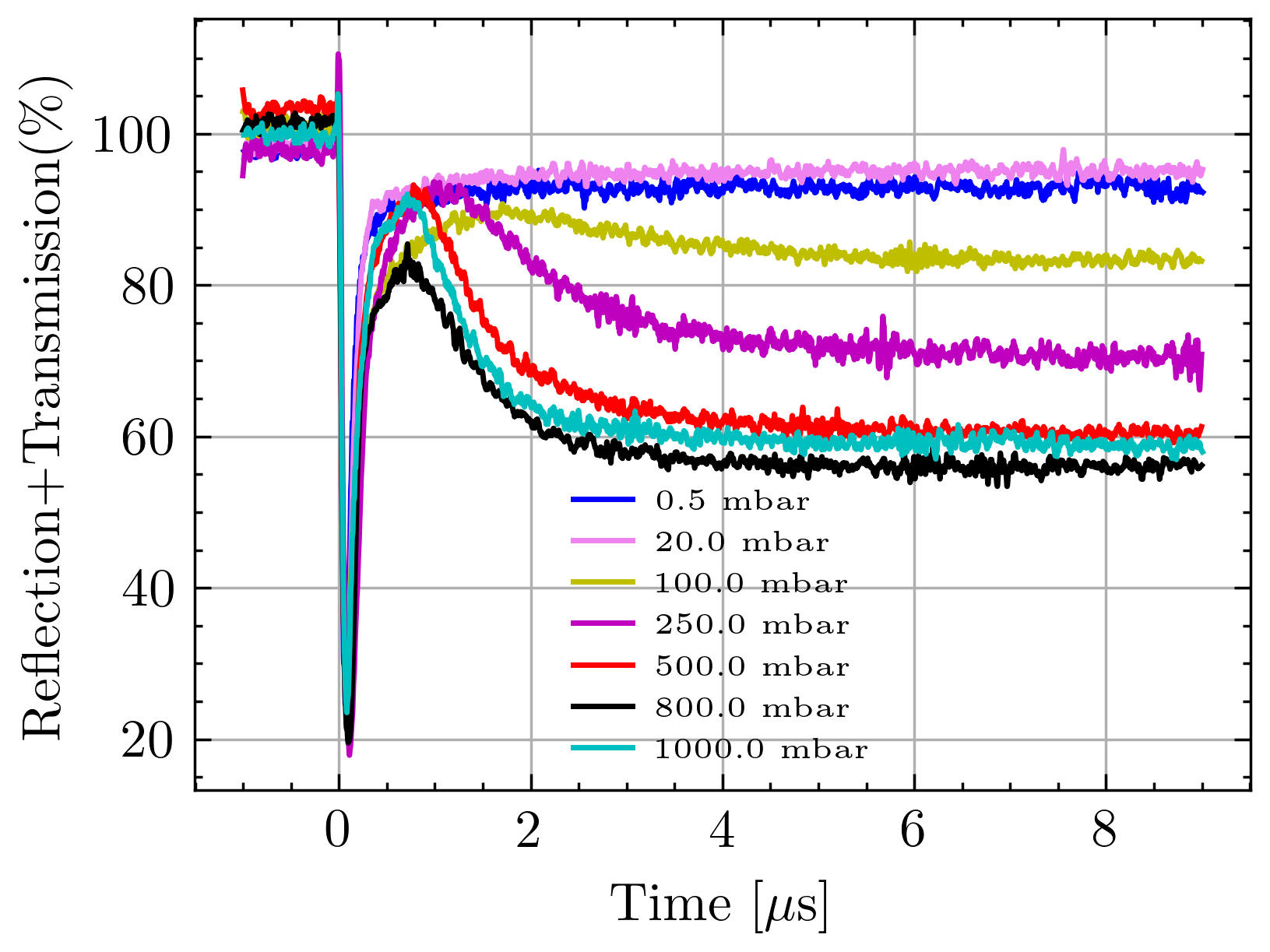}
		\caption{Sum of reflection and transmission at a fluence of 5.0 $\text{ J/cm}^{2}$ under different ambient pressure}
		\label{fig:absorption_20mj_var_P}
	\end{figure}
	
\begin{figure}[hbtp]
		\centering
		\includegraphics[width=1\textwidth]{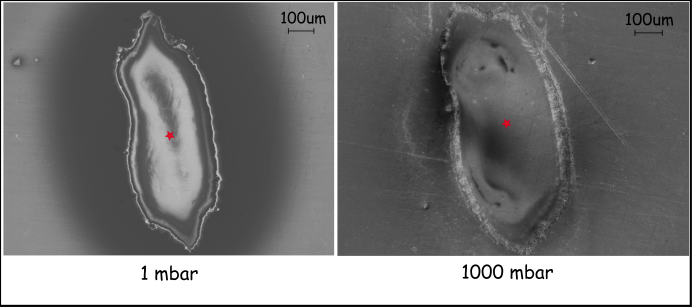}
		\caption{SEM Images of the ablated area during the initial pulse for 1mbar and 1000mbar pressure conditions laser fluence of 5.0 $\text{ J/cm}^{2}$ * represents the region where micro-Raman spectrum was performed)}.
		\label{fig:ba_sem_comb}
	\end{figure}

	It will be of interest to check the combined behaviour of transmission and reflection from the ablation site during the interaction of the pump laser with the film. Figure~\ref{fig:absorption_20mj_var_P} illustrates the behaviours of transmission and reflection from different background pressures for a fluence of $5.0 \text{ J/cm}^{2}$. It is anticipated that the sum of transmission and reflection will be unity if absorption and scattering are not there. However, figure \ref{fig:absorption_20mj_var_P} illustrates a pronounced decrease in the sum at approximately 100 nanoseconds after ablation, followed by a swift recovery. After the dip at 100 nanoseconds, the sum remains at approximately 100\% for lower background pressures, as anticipated. On the other hand, for higher pressures, the sum again reduces, indicating a loss in reflection from the redeposited film. This loss in reflection could be a consequence of diffused reflection of the re-deposited film, which may be attributed to its morphology. 
	As the results observed with varying background pressures provided further evidence in support of the argument for re-deposition, the ablated spot was subjected to further characterisation using Micro Raman and SEM techniques. Figure \ref{fig:ba_sem_comb} displays the SEM of the ablated crater for different background pressures and pump laser fluences of 5.0 $\text{ J/cm}^{2}$. The SEM images clearly illustrate the differences in surface morphology within the ablated spot.
	At lower pressures, the image reveals a non-conducting surface with electron accumulation, indicating the complete removal of the aluminum film, as is evidenced by the presence of a white patch. However, at higher pressures, such as 500 mbar and 1000 mbar, the SEM image shows a uniform surface similar to the surrounding area, confirming the presence of aluminium films at the ablated spot. This confirms the re-deposition at high background pressure. 
	
\begin{figure}[hbtp]
		\centering
		\centering
		\includegraphics[width=1\textwidth]{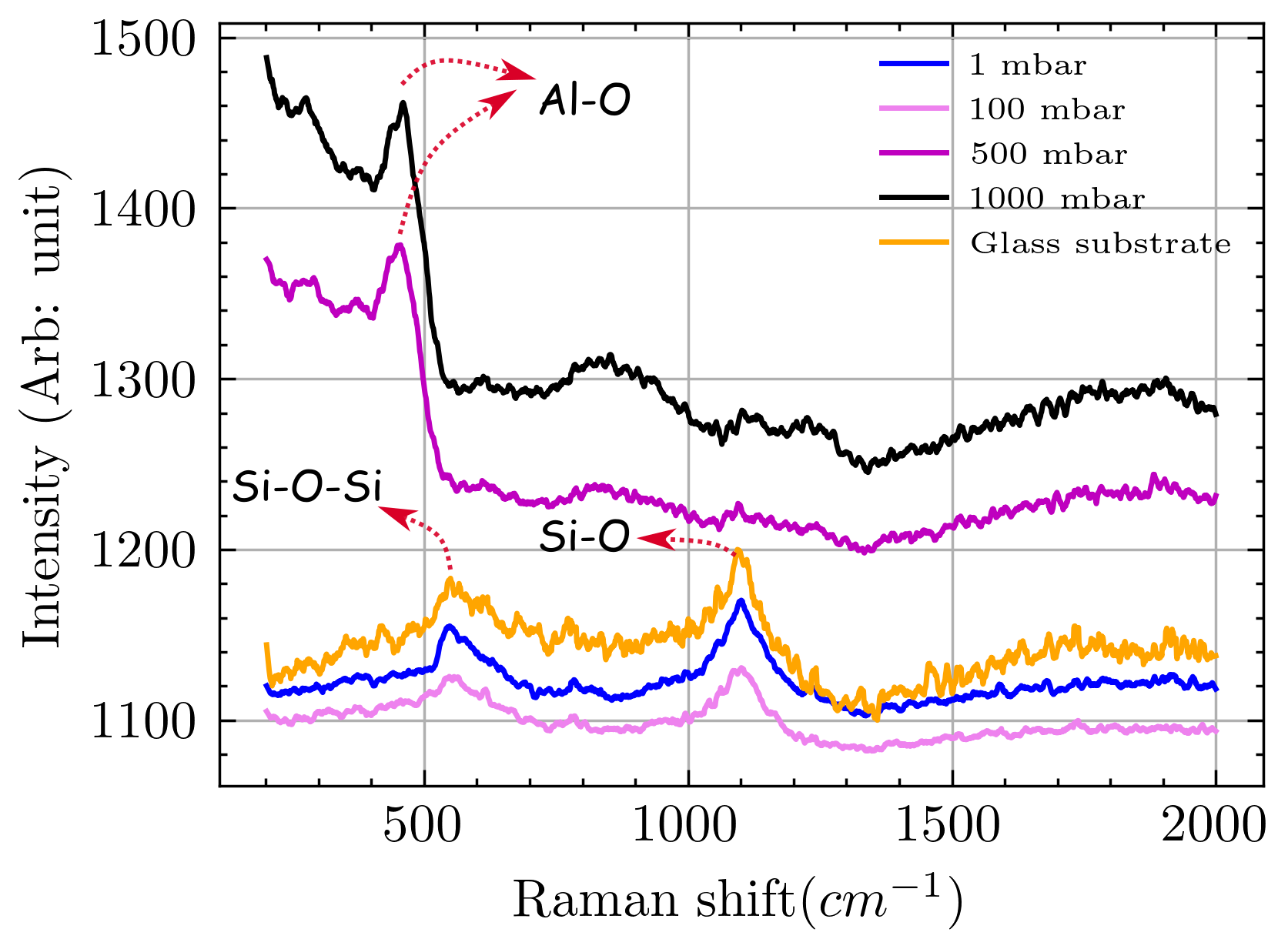}
		\caption{Raman spectroscopy snalysis of crater formation at a fluence of 5.0$\text{ J/cm}^{2}$ undere different ambient pressures. For clarity the micro-Raman spectrum of glass substrate is also recorded.}
		\label{fig:raman_20mj.png}
	\end{figure}

	Figure~\ref{fig:raman_20mj.png} shows the micro Raman spectra acquired from the ablation spots of film at different background pressures.  At lower pressures the Raman spectra shows the typical peaks corresponding to the glass substrate\cite{deschamps2011soda}. However, for background pressures of 500 and 1000 mbar the spectra do not show the peaks of glass substrate instead a peak at 480 $cm^{-1}$\cite{dienifer_raman} indicative of the presence of aluminium oxides appears. Hence it can be safely assumed that the re-deposition at higher background pressure occurs in the form of aluminium oxide. In short, the pump probe experimental data and the analysis of the spot using SEM and micro Raman reveal the re deposition of the plume back at the film substrate in the form of aluminium Oxide, which further shows a dependence with background pressure. Additionally, atomic force microscopy (AFM) analysis was conducted at the site of ablation to investigate the influence of background pressure. Similar to the SEM and micro-Raman observations, the AFM study also indicates that a significant amount of film ($\approx 150 $ nm) remains at the spot of ablation at atmospheric pressure in comparison to the ablation at 1 mbar of background pressure for the fluence of 5 $\text{ J/cm}^2$. This suggests that a significant amount of material is redeposited to the location of ablation at atmospheric pressure.
	
	A substantial transmission through the ablation site, even in the presence of a moderately thick film, eliminates the possibility of pure aluminum coatings, as such coatings are expected to exhibit high reflectivity and absorbance\cite{Marla_2014}. However, as previously reported, aluminum oxides exhibit higher transmission and lower reflection than pure aluminum \cite{Koushki_aluminum, Zhao_aluminium, dobrzanski2015surface}.  The micro-Raman analysis indicates that the redeposited spots indicate the formation of aluminum oxide at elevated background pressure. Therefore, the observed transmission from the redeposited spots is reasonable. Further studies are necessary to elucidate exact nature of the redeposition of this oxide layer during laser ablation.
	
	\section{Conclusion}\label{sec:conclusion}
	
	\label{sec:conclusion}
	
	The present work demonstrates an innovative experimental approach for the study of laser surface interaction using a conventional pump-probe setup. The study reveals a number of processes involved in laser surface interaction, e.g. fast-melting and re-solidification, light interaction with high-density plasma, plasma interaction with the surrounding medium, and the effect of the surrounding medium on the re-deposition of the ablated mass back to the substrate. It provides evidence in support of some of the recent simulations that predict re-deposition. The experimental findings from the pump-probe experiments are further validated through morphological studies, such as scanning electron microscopy (SEM) and micro-Raman spectroscopy. The studies demonstrate that at higher background pressure, irrespective of the laser fluence used, substantial amount of film remains present. The films initially melt and then a portion in contact with the substrate immediately re-solidifies, while a larger portion is re-deposited due to the short stopping distance at higher background pressures. The interesting dynamics of the probe signal exhibit characteristics may warrant further experimental and theoretical investigations for a comprehensice picture. Nonetheless present study using evolution of reflectance and transmission bring out some interesting features in the ablation of thin film.
	
	We believe that findings should be of greater importance in the context of potential applications, such as insitu cleaning of viewing windows in large experimental facilities, such as tokamaks. In such facilities, the viewports may become contaminated with different elements, with a decrease in their transmission and affecting the diagnostics. Further it may also find applications in pulsed laser deposition(PLD) and laser induced forward transfer(LIFT), laser induced surface structuring/modification.
	
	\section*{Data availability} 
	
	The data that supports the observations of this study are available from the corresponding author upon reasonable request.
	\section*{Acknowledgment} 
	The authors acknowledge the help received from K. M.Yatheendran, Raman Research Institute, Banglore, India for assistance with SEM measurements, Dr. Balasubramanian C and Ms. Tarundeep Kaur Lamba, Institute for Plasma Research, for AFM analysis of the sample.
	\section{Reference}

\end{document}